\documentclass[journal]{IEEEtran}
\usepackage{amsmath,amsfonts}
\usepackage{amssymb}
\usepackage{algorithmic}
\usepackage{multirow}
\usepackage{cite}
\usepackage{array}
\usepackage[caption=false,font=normalsize,labelfont=sf,textfont=sf]{subfig}
\usepackage{textcomp}
\usepackage{stfloats}
\usepackage{url}
\usepackage{verbatim}
\usepackage{graphicx}
\usepackage{booktabs}
\usepackage[urlcolor=blue]{hyperref}
\makeatletter
\newcommand*\bigcdot{\mathpalette\bigcdot@{.5}}
\newcommand*\bigcdot@[2]{\mathbin{\vcenter{\hbox{\scalebox{#2}{$\m@th#1\bullet$}}}}}
\makeatother
\ifCLASSINFOpdf
\else
\fi
\def\BibTeX{{\rm B\kern-.05em{\sc i\kern-.025em b}\kern-.08em
    T\kern-.1667em\lower.7ex\hbox{E}\kern-.125emX}}
\usepackage{balance}

\begin{document}
\title{LeVo 2: Stable and Melodious Song Generation via Hierarchical Representation Modeling and Progressive Post-Training}
\author{Shun Lei,
        Huaicheng Zhang,
        Dapeng Wu,
        Yaoxun Xu,
        Lishi Zuo,
        Wei Tan,
        Hangting Chen,
        Guangzheng Li,
        Jianwei Yu,
        Zhiyong Wu,
        Dong Yu,~\IEEEmembership{Fellow,~IEEE}
\thanks{Corresponding author: Zhiyong Wu and Dong Yu.}
\thanks{Shun Lei, Dapeng Wu, Yaoxun Xu, and Zhiyong Wu are with Tsinghua-CUHK Joint Research Center for Media Sciences, Technologies and Systems, Shenzhen International Graduate School, Tsinghua University, Shenzhen, China (e-mail: leis21@mails.tsinghua.edu.cn; wdp24@mails.tsinghua.edu.cn; xuyx22@mails.tsinghua.edu.cn; zywu@sz.tsinghua.edu.cn). This work was conducted when Shun Lei was an intern at Tencent.}
\thanks{Wei Tan, Hangting Chen, Guangzheng Li, Jianwei Yu, and Dong Yu are with Tencent, Shenzhen, China (e-mail: waytan@tencent.com; chenhangting17@mails.ucas.ac.cn; philipgzli@tencent.com; tomasyu@foxmail.com; dongyu@ieee.org).}
\thanks{Huaicheng Zhang is with Wuhan University, Wuhan, China (e-mail: zhuaicheng@whu.edu.cn).}
\thanks{Lishi Zuo is with Hong Kong Polytechnic University, Hong Kong, China (e-mail: lishi.zuo@connect.polyu.hk).}}

\markboth{Journal of \LaTeX\ Class Files,~Vol.~18, No.~9, September~2020}%
{Shell \MakeLowercase{\textit{Shun Lei et al.}}: Stable and Melodious Song Generation via Hierarchical Representation Modeling and Progressive Post-Training}
\maketitle

\begin{abstract}
Full-length song generation must preserve coherence and musicality, render detailed vocal and accompaniment acoustics, and follow lyrics and prompts.
Existing language model-based systems face a structural trade-off: mixed-token modeling preserves vocal-instrument coordination but obscures track-specific details, whereas dual-track prediction improves acoustics but requires longer sequences and weakens global planning.
We present LeVo 2, a hybrid LLM-Diffusion framework for controllable full-length song generation.
LeVo 2 formulates this trade-off as hierarchical modeling: LeLM first predicts mixed tokens for semantic planning, then predicts vocal and accompaniment tokens in parallel for track-specific refinement, while a diffusion-based Music Codec reconstructs full-length waveforms.
A central contribution of this extended version is an aesthetics-guided training schedule for alignment.
During pre-training, an automated music aesthetic evaluation framework assigns musicality-tier conditions to large-scale data, providing musicality priors before preference alignment.
Progressive post-training applies SFT, large-scale offline DPO, and closed-loop semi-online DPO to separately improve generation quality, controllability, and musicality.
Modular extension then trains the Track-Specific LM for acoustic refinement while preserving the aligned semantic planner.
This schedule separates musicality learning, controllability alignment, and acoustic refinement, mitigating optimization conflict and the limitations of static offline preference pairs.
Expert listening tests and objective evaluations show that LeVo 2 outperforms open-source baselines across six subjective dimensions, and approaches leading commercial systems on several listening metrics.
Ablations validate the effects of the training strategy, aesthetics guidance, scaling, and hierarchical architecture.
\end{abstract}


\IEEEpeerreviewmaketitle
\ifCLASSOPTIONcaptionsoff
  \newpage
\fi

\section{Introduction}
\label{sec:introduction}
Music serves as a fundamental pillar of human culture, with songs standing out by uniquely blending expressive vocals and rich instrumental accompaniment to convey the emotions and thoughts of creators.
Driven by rapid advancements in Artificial Intelligence Generated Content (AIGC), automated music generation has become an active research topic in both academia and industry.
While remarkable progress has been made in specific sub-domains---including symbolic music generation \cite{ismir/YangCY17, dong2018musegan, roberts2018hierarchical, yu2022museformer}, instrumental music generation \cite{musiclm, mousai, musicgen, melody, majumder2024tango}, and singing voice synthesis \cite{chen2020hifisinger, huang2022singgan, zhang2022visinger, liu2022diffsinger, hong2023unisinger}---automatically generating a complete song remains a substantial challenge.
It requires generative models to simultaneously synthesize and seamlessly integrate expressive vocals and high-fidelity instrumental tracks while preserving musicality and following instructions.
This task is further complicated when generating full-length songs, where maintaining lyric alignment and long-term structural coherence become more demanding.

Early attempts in song generation, such as Jukebox \cite{dhariwal2020jukebox}, utilized language models (LMs) to predict discrete codes---referred to as ``mixed tokens''---extracted from the combined audio of vocals and accompaniment.
Subsequent studies \cite{lei2024songcreator, lam2025analyzable, yang2025songbloom, jiang2026muse, yang2026heartmula} further optimized LM architectures and modeling paradigms to enhance overall musicality.
Nevertheless, mixed token-based approaches exhibit critical limitations. 
Specifically, the restricted discrete vocabulary and the mutual acoustic masking of audio tracks obscure fine-grained details and hinder high-fidelity generation.
To address this, YuE \cite{yuan2025yue} introduces a dual-track token prediction strategy to separately model vocal and accompaniment tracks.
SongGen \cite{liu2025songgen} extends this idea and shows that the interleaved token prediction strategy effectively reduces interference between different token types. 
While these methods produce high-quality individual tracks, their independent prediction mechanisms often struggle to maintain vocal-instrument harmony. 
Furthermore, the interleaved paradigm drastically increases the sequence length, severely limiting model scalability and exacerbating the difficulty of maintaining musicality and following instructions.
Concurrently, diffusion-based systems, exemplified by the DiffRhythm \cite{ning2025diffrhythm, chen2025diffrhythm+, jiang2025diffrhythm} and ACE-Step \cite{gong2025ace, gong2026ace} series, avoid discrete tokens but encounter challenges in accurately capturing temporal phoneme-acoustic alignments and maintaining long-term structural consistency across full-length songs.
Recent industry song generation tools \cite{suno, udio, mureka, minimax} have also emerged, but none of these systems has disclosed its methodology or training strategy.

Beyond architectural limitations, the song generation community has long been constrained by the scarcity of high-quality data and noisy annotations.
On the one hand, while models trained on such datasets can successfully fit the general music distribution, they lack the musicality priors required to align the generated songs with human preferences.
On the other hand, unreliable data annotations impair the model's instruction following capabilities, leading to lyrical hallucinations and unstable controllability. 
To mitigate these issues, recent studies have turned to preference alignment.
DiffRhythm+ \cite{chen2025diffrhythm+} employs Direct Preference Optimization (DPO) to enhance musicality, while LeVo \cite{leilevo}, DiffRhythm 2 \cite{jiang2025diffrhythm}, and HeartMuLa \cite{yang2026heartmula} further explore multi-preference alignment strategies to simultaneously improve both musicality and instruction-following abilities.
Nevertheless, these existing approaches exhibit two limitations. 
First, simultaneously optimizing multiple preference objectives often induces gradient conflicts.
The mutual interference between different objectives may lead to an averaging effect, which restricts the upper bound of each individual capability.
Second, these methods rely exclusively on static, offline paired datasets.
This not only restricts the model's performance upper bound within the capabilities of the pre-trained models used for dataset construction, but also renders the alignment process susceptible to reward hacking.

LeVo 2 treats the trade-off between global song coherence and track-level acoustic detail as a hierarchical modeling problem.
Its core component, LeLM, separates global semantic planning from track-specific acoustic refinement through two coordinated token prediction tasks.
The Mixed Semantic LM predicts mixed tokens to capture song-level musical structure, including melody, rhythm, tempo, and vocal-instrument coordination.
Conditioned on the hidden states of this global semantic planner, the Track-Specific LM predicts vocal and accompaniment tokens in parallel, recovering fine-grained acoustic details while avoiding the sequence-length expansion caused by interleaved dual-track prediction.
The diffusion-based Music Codec then reconstructs the predicted tokens into full-length song waveforms.

The main training contribution of LeVo 2 is a decoupled, aesthetics-guided alignment strategy for full-length song generation.
Training a controllable song generator requires different learning signals for musicality, lyric alignment, prompt consistency, and track-level acoustic refinement.
LeVo 2 assigns these objectives to separate stages instead of optimizing them under a single preference objective.
During pre-training, LeVo 2 scales model capacity and training data, includes pure instrumental music to strengthen accompaniment modeling, and uses an automated music aesthetic evaluation framework to assign musicality-tier conditions.
These conditions provide musicality priors before preference alignment.
During progressive post-training, SFT first narrows the generation distribution toward high-quality songs, large-scale offline DPO then improves lyric alignment and prompt consistency, and closed-loop semi-online DPO further improves musicality using samples from the evolving policy.
This schedule mitigates optimization conflict and the limitations of static offline preference pairs.
Finally, modular extension trains the Track-Specific LM to refine vocal and accompaniment acoustics while preserving the aligned global semantic planner.

Extensive experiments demonstrate the superiority of our approach. 
Through a large-scale expert subjective evaluation, we comprehensively assessed LeVo 2 across six core dimensions: Overall Musicality, Melody, Arrangement, Instrumental Sound Quality, Vocal Sound Quality, and Structure. 
LeVo 2 significantly outperforms all open-source baselines across all dimensions and approaches leading closed-source systems on several listening metrics. 
Objective evaluations further validate its capabilities in lyric alignment and versatile control over genres, emotions, and instruments.
Audio examples, inference code, and full model weights are released\footnote{Audio examples are available at \href{https://levo-demo.github.io/levo_v2_demo/}{https://levo-demo.github.io/levo\_v2\_demo/}. Inference code and full model weights are available at \href{https://github.com/levo-demo/LeVo}{https://github.com/levo-demo/LeVo}.} to facilitate future research.

This paper builds upon our previous work \cite{leilevo}, which introduced the overall architecture of LeVo 2. 
The additional contributions of this extended version are summarized as follows:

\begin{itemize}
    \item We propose an aesthetics-guided three-stage training paradigm (pre-training, progressive post-training, and modular extension). Guided by an automated music aesthetic evaluation framework, this approach continuously injects musicality priors into the model and introduces a musicality-aware Classifier-Free Guidance (CFG) strategy that significantly enhances inference quality.
    \item We introduce a decoupled multi-stage progressive post-training scheme consisting of SFT, large-scale offline DPO, and closed-loop semi-online DPO. By applying this alignment to the Mixed Semantic LM, we resolve the complex conflicts among controllability, stability, and musicality, alleviating lyrical hallucinations while maximizing artistic expressiveness.
    \item We investigate the scaling effects within LeVo architecture and the incorporation of pure instrumental data, which improves the modeling of complex background accompaniments and arrangement.
    \item We conduct comprehensive large-scale subjective and objective evaluations against commercial systems and open-source models. The results demonstrate steady performance improvements achieved at each training stage, while extensive ablation studies validate the effectiveness of our core technical designs.
\end{itemize}


\section{Related Work} \label{sec:relatedwork}
\subsection{Music Generation}
Early efforts in music generation primarily focused on symbolic music generation \cite{dong2018musegan,wei23c_interspeech}.
However, symbolic music inherently lacks the expressive nuances, such as timbre, articulation, and performance dynamics, which are crucial for high-fidelity and melodious music synthesis.
Driven by the scaling behaviors and reasoning capabilities of Large Language Models (LLMs) \cite{lamda, achiam2023gpt, touvron2023llama}, several studies \cite{audiolm, musiclm, musicgen} have shifted toward end-to-end music generation.
These approaches typically discretize music into token sequences via Vector Quantised-Variational Autoencoders (VQ-VAE) \cite{van2017neural} or Residual Vector Quantization (RVQ) \cite{zeghidour2021soundstream, kumar2023high} for language model processing. 
Despite their success, the performance of these models is often constrained by quantization loss.

To address this, diffusion models have been adopted for modeling continuous representations, demonstrating effectiveness in high-fidelity synthesis \cite{peebles2023scalable, rombach2022high}.
However, these methods face significant computational challenges in long-context generation as sequence length and model size increase, making it difficult to capture temporal phoneme-acoustic alignments and maintain long-term structural consistency across full-length music.
Recently, hybrid architectures such as MeLoDy \cite{melody} and AudioLDM 2 \cite{liu2023audioldm} have combined LMs with diffusion models, achieving performance increases by balancing semantic modeling with acoustic fidelity.
However, these works are limited to instrumental music generation and struggle with the composition of songs, where vocals and accompaniment must remain harmonized.
LeVo 2 extends this hybrid paradigm by utilizing a diffusion-based Music Codec as the high-fidelity renderer, integrated with a hierarchical LeLM. 
By explicitly dividing modeling responsibilities into a Mixed Semantic LM for global semantic and a Track-Specific LM for acoustic details, LeVo 2 is uniquely designed to capture the structural and acoustic complexities of song generation.

\subsection{Song Generation}
Song generation aims to synthesize expressive vocals integrated with rich accompaniment.
Pioneering works like Jukebox \cite{dhariwal2020jukebox} utilized cascaded Transformers to model mixed discrete tokens extracted from songs.
To enhance the musicality, SongCreator \cite{lei2024songcreator} introduced a dual-sequence language model to capture the relationship between vocals and accompaniment.
MusiCot \cite{lam2025analyzable} predicts coarse-grained style representations to guide the generation of audio tokens.
Furthermore, SongBloom \cite{yang2025songbloom} and HeartMuLa \cite{yang2026heartmula} attempted to enhance sound quality by modeling more complex acoustic representations.
However, these mixed token-based approaches remain constrained by restricted vocabularies and mutual acoustic masking between vocals and accompaniment, which inherently limits audio fidelity.

To address these issues, Melodist \cite{hong2024text} and MelodyLM \cite{li2024accompanied} utilize a multi-stage process to sequentially produce vocals and accompaniment.
YuE \cite{yuan2025yue} and SongGen \cite{liu2025songgen} attempt to operate on sequences consisting of vocal tokens and accompaniment tokens. 
While these methods improve individual track quality, their independent prediction mechanisms often struggle with vocal-instrument harmony.
Furthermore, the interleaved paradigm increases the sequence length, severely limiting model scalability and increasing the difficulty of maintaining musicality and following instructions. 
Alternatively, diffusion-based systems like DiffRhythm \cite{ning2025diffrhythm, chen2025diffrhythm+, jiang2025diffrhythm} and ACE-Step \cite{gong2025ace, gong2026ace} series avoid discretization but face challenges in phoneme-acoustic alignment and maintaining long-term consistency. 
Though commercial systems \cite{suno, udio, mureka, minimax} show strong performance, their methodologies remain proprietary.

LeVo 2 advances this field via a hierarchical representation modeling mechanism. 
Unlike previous works, it explicitly assigns mixed tokens to the Mixed Semantic LM to ensure global harmony, and dual-track tokens to the Track-Specific LM to enhance acoustic fidelity. 
This hierarchical architecture not only circumvents the sequence length explosion in interleaved models but also prevents mutual interference between different token types by completely decoupling their modeling.

\subsection{Post-training of music generation models}
Reinforcement Learning (RL) techniques have recently been introduced to music generation for preference alignment.
BATON \cite{liao2024baton} and MusicRL \cite{cideron2024musicrl} utilize reward models trained based on human feedback to guide diffusion and language models, respectively.
Tango 2 \cite{majumder2024tango} employs DPO with semi-automated preference datasets to specifically enhance instruction-following capabilities, while DiffRhythm+ \cite{chen2025diffrhythm+} refines overall musicality via DPO. 
However, these efforts primarily focus on single-dimension optimization.

HeartMuLa \cite{yang2026heartmula} constructs specialized datasets for various preferences and performs mixed training to align multiple preferences simultaneously. 
LeVo \cite{leilevo} introduces an interpolation-based method to merge different optimization objectives within the DPO framework. 
To further mitigate the complexity of multi-preference learning, DiffRhythm 2 \cite{jiang2025diffrhythm} proposes a cross-pair preference optimization strategy to provide a more stable gradient signal for multiple musical dimensions.

Despite these advancements, these strategies often suffer from gradient conflicts and an averaging effect. Specifically, when mutual interference exists between the multiple dimensions, the simultaneous pursuit of these objectives leads to suboptimal trade-offs, thereby restricting the performance upper bound of each individual capability. 
Furthermore, their reliance on static, offline datasets not only limits the model's performance within the capacity of the data generators but also renders the training process susceptible to reward hacking. 
To circumvent these issues, LeVo 2 introduces an  aesthetics-guided, decoupled multi-stage progressive post-training scheme.
By sequentially addressing controllability and musicality through SFT, large-scale offline DPO, and closed-loop semi-online DPO, our approach eliminates mutual interference and iteratively pushes the boundaries of artistic expressiveness beyond the constraints of static data.

\begin{figure*}[!t]
     \centering
     \includegraphics[width=\textwidth]{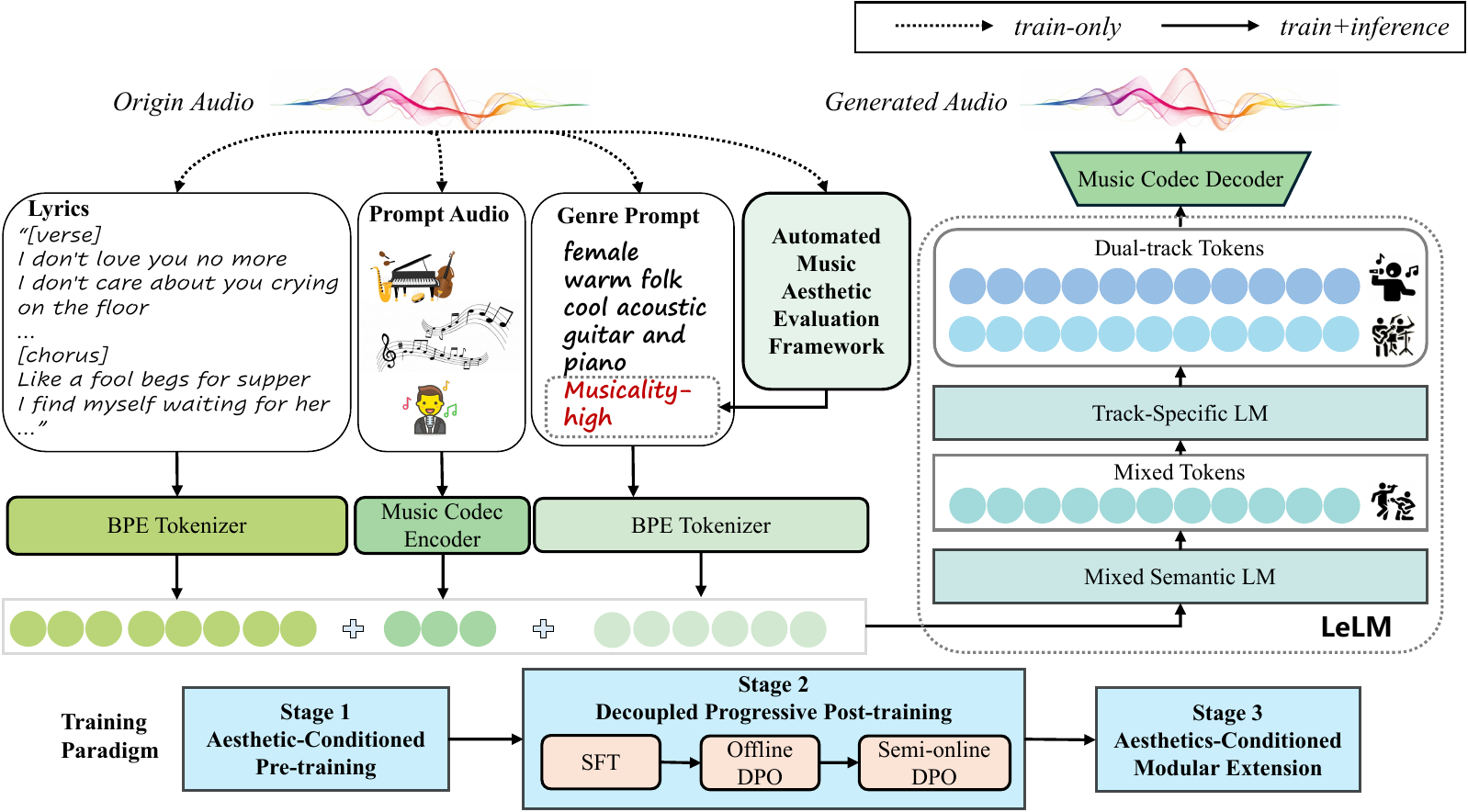}
     \caption{Overview of LeVo 2. LeLM performs hierarchical semantic planning over mixed and dual-track tokens, while the diffusion-based Music Codec reconstructs the generated tokens into full-length song audio. The bottom panel summarizes the training paradigm: pre-training establishes global semantic planning, decoupled progressive post-training improves controllability and musicality through SFT, large-scale offline DPO, and closed-loop semi-online DPO, and modular extension refines track-specific acoustics.}
     \label{fig:overall}
\end{figure*}
\section{Method} \label{sec:method}
\subsection{Overview} \label{sec:overview}
Let $\mathbf{x}\in\mathcal{X}$ represent a song audio, and $\mathcal{C}$ denote the set of conditions (e.g., lyrics, audio prompt, and text description).
The goal of song generation is to learn a mapping $f: \mathcal{C} \mapsto \mathcal{X}$ that synthesizes high-fidelity audio conditioned on $\mathbf{C} \in \mathcal{C}$.
Since direct end-to-end synthesis remains a challenge, we follow the paradigm of language model-based music generation \cite{melody, lei2024songcreator, yuan2025yue} and bridge this gap by introducing discrete token sequences $\mathbf{S} = (S_1, \ldots, S_N)$ to capture the essential structural and semantic information of the song.

As illustrated in Figure \ref{fig:overall}, LeVo 2 adopts a hybrid LLM-Diffusion architecture consisting of two core components: LeLM, which captures the semantics of multiple tracks, and a diffusion-based Music Codec, which serves as the high-fidelity renderer.
To extract prediction targets for training, the Music Codec encoder discretizes the song audio into mixed tokens ($\mathbf{S}_m \in \mathcal{S}_m$), vocal tokens ($\mathbf{S}_v \in \mathcal{S}_v$), and accompaniment tokens ($\mathbf{S}_a \in \mathcal{S}_a$).
The LeLM is designed with a hierarchical architecture utilizing two LMs to model these token streams in parallel. 
Specifically, given the prefix context $\mathbf{C}$, LeLM simultaneously predicts mixed tokens to provide global semantic guidance and dual-track tokens ($\mathbf{S}_v, \mathbf{S}_a$) to capture track-specific nuances. 
Finally, the diffusion-based decoder of the Music Codec reconstructs the predicted dual-track tokens into high-quality music audio.

To further address the challenges of token interference, musicality, and instruction following, we implement a three-stage training paradigm guided by an automated music aesthetic evaluation framework, together with a decoupled multi-stage progressive post-training strategy. 
The following subsections provide detailed descriptions of the LeVo 2 architecture and its progressive optimization process.

\subsection{Hierarchical Language Modeling} 
\begin{figure*}
     \centering
     \includegraphics[width=\textwidth]{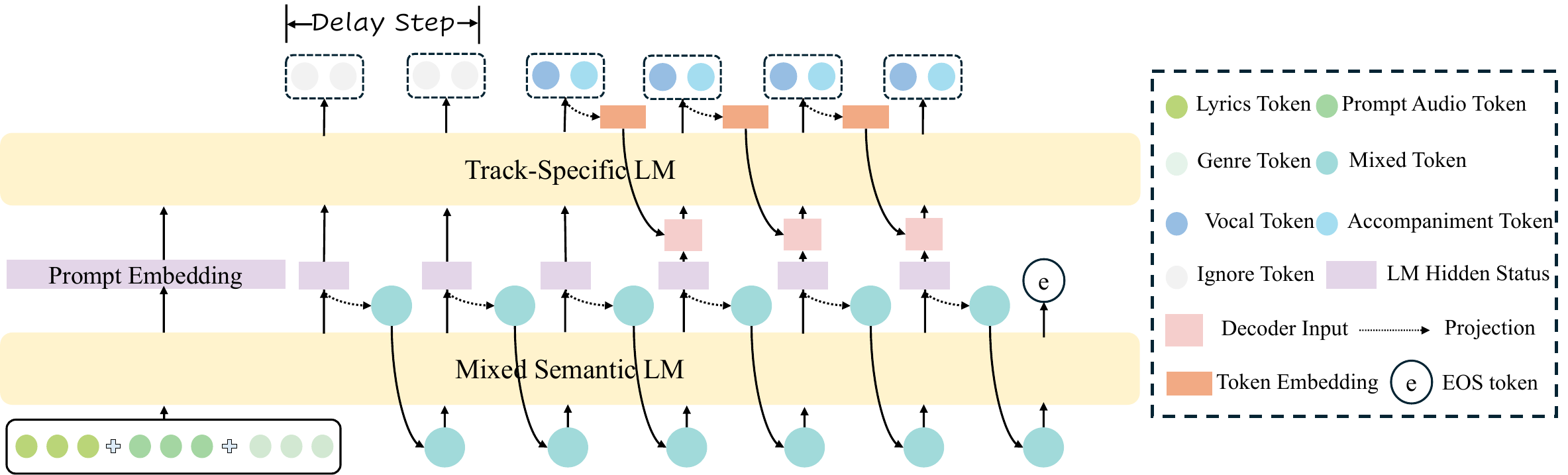}
     \caption{
     The architecture of LeLM, which consists of a Mixed Semantic LM for global semantic modeling and a Track-Specific LM for parallel track refinement.}
     \label{fig:lm}
\end{figure*}
Although modeling dual-track tokens offers significant advantages in sound quality, maintaining vocal-instrument harmony remains a challenge \cite{yuan2025yue, liu2025songgen}.
While the interleaving pattern can reduce cross-token interference compared to naive parallel patterns, its substantial increase in sequence length limits performance for long-context generation.
To address this, LeVo 2 introduces a hierarchical parallel modeling approach.
By decoupling global orchestration from track-specific details and modeling them in parallel, this design eliminates cross-token interference without lengthening the sequence, thus simultaneously optimizing vocal-instrument harmony and sound quality.
As illustrated in Figure \ref{fig:lm}, this hierarchical architecture is achieved through 
the collaboration of two specialized components: the Mixed Semantic LM and the Track-Specific LM.

The Mixed Semantic LM serves as the framework's global semantic planner.
It utilizes a decoder-only Transformer architecture, which has been widely adopted in previous works on music generation \cite{melody, musiclm, yuan2025yue}.
It focuses on the next-token prediction task for mixed tokens, which acts as a pseudo-chain-of-thought to capture high-level structural information such as melody, rhythm, and tempo to provide a global semantic blueprint that ensures intrinsic vocal-instrument harmony.
This task can be formulated as:
\begin{equation}
p(\mathbf{S}_m|\mathbf{C};\boldsymbol\theta) = \prod_{t=0}^T p(\mathbf{S}_{m,t}|\mathbf{S}_{m,<t},\mathbf{C};\boldsymbol\theta)
\end{equation}
where $\mathbf{C}$ denotes conditions, including lyrics, optional text descriptions, and optional audio prompts.

To facilitate high-quality song generation, we employ the Track-Specific LM, a lightweight module designed to refine the global semantic guidance into finer acoustic details.
It predicts vocal tokens ($\mathbf{S}_v$) and accompaniment tokens ($\mathbf{S}_a$) in parallel, conditioned on the representations from the Mixed Semantic LM. 
Since the hidden states of the Mixed Semantic LM contain richer information than the discrete mixed tokens, we pass these hidden states directly to the Track-Specific LM.
Specifically, at each time step $t$, the embeddings of the predicted vocal and accompaniment tokens from the previous time step, along with the hidden states of the Mixed Semantic LM's last layer, are concatenated to form the input of the Track-Specific LM. 
Two groups of linear heads are then employed to project the output into the respective discrete token spaces.

Moreover, to provide more comprehensive contextual information for specific-track modeling, we introduce a delay pattern mechanism. 
This means that when predicting the dual-track tokens at the $t$-th time step, the model considers the hidden states up to $t+k$, where $k$ represents the number of delay steps.
This process can be simply represented as:
\begin{small}
\begin{equation}
\begin{split}
    p(\mathbf{S}_v,\mathbf{S}_a|\mathbf{C};\boldsymbol\theta) = & \prod_{t=0}^{T-k} p(\mathbf{S}_{v,t},\mathbf{S}_{a,t}|\mathbf{S}_{v,<t},\mathbf{S}_{a,<t},\mathbf{S}_{m,<t+k},\mathbf{C};\boldsymbol\theta)\\
    & \prod_{t=T-k+1}^T p(\mathbf{S}_{v,t},\mathbf{S}_{a,t}|\mathbf{S}_{v,<t},\mathbf{S}_{a,<t},\mathbf{S}_{m},\mathbf{C};\boldsymbol\theta)
\end{split}
\end{equation}
\end{small}

\subsection{Music Codec}
\begin{figure*}
     \centering
     \includegraphics[width=\textwidth]{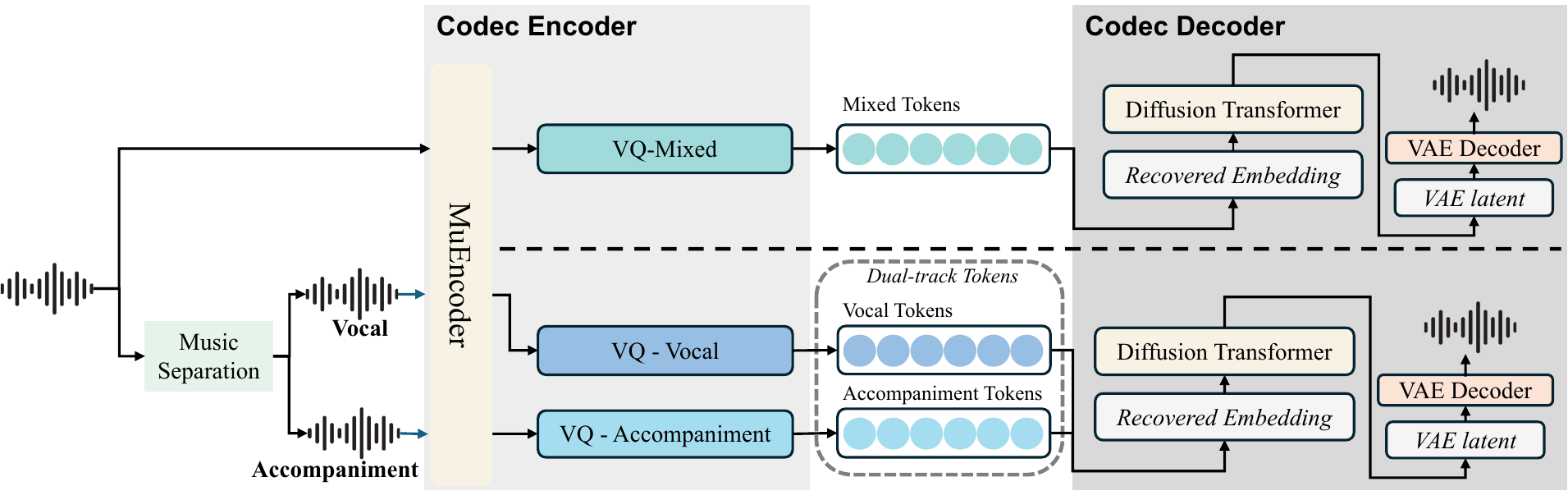}
     \caption{
     The framework of the Music Codec in LeVo 2. 
     }
     \label{fig:codec} 
\end{figure*}

Building upon MuCodec \cite{xu2024mucodec}---an efficient 48kHz music codec capable of operating at low bitrates---we design the Music Codec for LeVo 2, as shown in Figure \ref{fig:codec}.
The codec comprises an encoder for target discretization and a diffusion-based decoder for audio reconstruction.

The encoder is composed of the MuEncoder \cite{xu2024mucodec} and specialized Vector Quantizers (VQs): the MuEncoder extracts semantic representations, while the VQs discretize these representations into tokens.
As the central intermediary in LeVo 2, these tokens serve not only as the prediction targets for LeLM but also play a pivotal role in achieving high-fidelity music reconstruction by acting as inputs for the decoder.
The codec decoder comprises a diffusion transformer and a Variational Autoencoder (VAE) decoder. 
The diffusion transformer reconstructs VAE latent features from the token-derived recovered embeddings, which the VAE decoder then converts directly into audio.
This approach reduces decoding cost compared with methods that rely on Mel-spectrograms as intermediate steps for audio reconstruction.

Music inherently exhibits a rich hierarchical structure, particularly in the complex interplay between vocals and accompaniment. 
To generate expressive vocals alongside diverse accompaniment, the codec is designed with the capability to process and reconstruct both mixed and dual-track audio streams, thereby supporting our hierarchical modeling paradigm.
For mixed semantic extraction, music is directly processed by the MuEncoder and a dedicated VQ (VQ-Mixed) to extract mixed tokens ($\mathbf{S}_m$). 
These tokens encapsulate information from both the vocals and the accompaniment, providing the training targets for the Mixed Semantic LM. 
The decoder can then use the recovered embeddings from these mixed tokens to reconstruct the VAE latent, finally restoring the original music.
To capture track-specific details, a pretrained music separation model is employed to separate the vocal and accompaniment tracks. 
Each track is then processed by the MuEncoder and its respective quantizer (VQ-Vocal and VQ-Accompaniment) to produce vocal tokens ($\mathbf{S}_v$) and accompaniment tokens ($\mathbf{S}_a$) for the Track-Specific LM. 
The decoder conditions on the recovered embeddings from both sets of dual-track tokens to reconstruct the joint VAE latent, and finally restores the high-fidelity music.

To facilitate efficient and stable modeling, we employ a two-stage training pipeline for the Music Codec. 
First, we pre-train a Variational Autoencoder (VAE) to compress the original 48kHz audio into a continuous latent space with a 25Hz frame rate and a 64-dimensional representation. 
Second, with both the pre-trained MuEncoder and the VAE frozen, we jointly train the VQ module and the decoder to discretize the semantic representations into tokens and reconstruct the corresponding VAE latents. 
To support our hierarchical paradigm, the modules corresponding to the mixed tokens and the dual-track tokens are trained independently.

\subsection{Aesthetics-Guided Three-Stage Training Paradigm}
To improve the performance of the hierarchical LeLM architecture, we propose an aesthetics-guided three-stage training paradigm: pre-training, progressive post-training, and modular extension.
This paradigm explicitly divides the responsibilities of two modules: the Mixed Semantic LM focuses on modeling global semantics to ensure overall musicality and vocal-instrument harmony, while the Track-Specific LM models the acoustic details of individual tracks.
Therefore, we first pre-train the Mixed Semantic LM to establish its global semantic planning capabilities. Subsequently, through progressive post-training, we further align its instruction-following and musicality priors, producing melodious and stable base compositions. 
Finally, the Track-Specific LM is trained to supplement fine-grained acoustic details to optimize local performance, improve sound quality, and enrich fine-grained nuances.

Unlike approaches that rely solely on scaling up data, our paradigm emphasizes the continuous injection of musicality prior knowledge.
To achieve this, we introduce an automated music aesthetic evaluation framework, which acts as an objective ``judge'' to guide data utilization and human preference optimization throughout the model's training cycle.

\subsubsection{Automated Music Aesthetic Evaluation Framework}
To quantify the quality of music, we develop a comprehensive aesthetic evaluation framework inspired by SongEval \cite{yao2025songeval}.
The framework leverages the representations extracted from the self-supervised learning model MuQ \cite{zhu2025muq} to predict multiple performance metrics of a song. 
It evaluates song generation through a cohesive hierarchy: first examining fundamental acoustic properties (Vocal and Accompaniment Sound Quality), then assessing dimensions that directly impact the subjective listening experience (Melody, Structure, Arrangement, and Mixing), and finally capturing the overall artistic resonance through a Musicality dimension.
Trained on a diverse dataset of 11,717 samples generated by top-tier models, this framework demonstrates exceptionally high correlation with human expert ratings. 

\subsubsection{Stage 1: Aesthetics-Conditioned Pre-training}
In the first stage, the Mixed Semantic LM is trained on large-scale music data to align the modalities between the various conditioning inputs and the mixed tokens. 
During this phase, the Track-Specific LM is frozen, allowing the model to focus solely on learning global semantics and vocal-instrument harmony.
To ensure robustness, both audio prompts and text descriptions undergo a 50\% random drop.
To facilitate effective learning and capture long-term structural dependencies, we employ a length-incremental learning strategy. 
The training begins with shorter audio segments and progressively increases the sequence length, ultimately training on full-length song data.

To inject musicality priors, all training data is evaluated by our aesthetic framework.
Based on the calculated average score and relative ranking within its category, we divide the data into five discrete musicality tiers using percentile thresholds: Top 5\%, 5\%-25\%, 25\%-50\%, 50\%-75\%, and the remaining bottom 25\%. 
The corresponding ``musicality tier tag'' is assigned as an additional condition.
This mechanism enables the model to learn core acoustic and semantic knowledge from massive, long-tail noisy data without compromising its generation capabilities. 
Furthermore, it mitigates the ``one-to-many'' mapping problem, reducing conditional distribution entropy to accelerate convergence, and unlocks the capability for musicality-aware Classifier-Free Guidance (CFG) during inference.

\subsubsection{Stage 2: Decoupled Multi-Stage Progressive Post-training}
Before expanding the model to capture dual-track acoustic details, we first align the Mixed Semantic LM with human preferences. 
In this stage, the aesthetic framework transitions from a conditioning tool to a data filter for SFT and a reward model for DPO. 
Semantic-level preference alignment reduces lyrical hallucinations and improves the model's musicality while lowering computational cost.
The detailed mechanisms of this progressive alignment process are elaborated in Section \ref{sec:dpo}.

\subsubsection{Stage 3: Aesthetics-Conditioned Modular Extension Training}
Building upon the aligned Mixed Semantic LM, we freeze its weights and exclusively train the Track-Specific LM to model dual-track tokens without disrupting the previously established semantic and musicality priors.
Consistent with Stage 1, the aesthetic tier tags are injected into the prefix context, ensuring that the parallel modeling of fine-grained nuances remains aligned with the high-quality musical standards.

During this stage, we initially employ a teacher-forcing scheme, utilizing mixed tokens extracted directly from real songs as inputs to predict the corresponding dual-track tokens. 
Once the training stabilizes, we introduce an acoustic augmentation strategy. Specifically, we reconstruct the input mixed tokens into low-quality audio using the Music Codec decoder with only 2 to 4 diffusion steps, and subsequently extract the mixed tokens from this degraded audio. 
This intentional corruption of acoustic details forces the Track-Specific LM to learn how to recover and supplement fine-grained acoustic nuances, optimize local track performance, and improve overall sound quality.

\subsection{Decoupled Multi-Stage Progressive Post-training} \label{sec:dpo}
Although the first two stages equip LeLM with base modeling capabilities, the model still faces challenges in musicality and instruction following due to the scarcity of high-quality data and noisy annotations. 
Therefore, multi-dimensional alignment with human preferences remains essential for stable and melodious song generation. 
Recent attempts at multi-preference alignment face two limitations: (1) simultaneously optimizing diverse objectives (e.g., objective controllability and subjective musicality) often induces gradient conflicts, leading to an averaging effect that restricts the upper bound of each capability; and (2) relying on static offline datasets caps the model's potential and increases susceptibility to reward hacking.
We use a decoupled multi-stage progressive post-training scheme for this alignment. It separates the optimization objectives into three phases: Supervised Fine-Tuning (SFT) narrows the data distribution and establishes a high-quality generation baseline, large-scale offline DPO improves controllability and reduces lyrical hallucinations, and closed-loop semi-online DPO improves artistic expressiveness.

\subsubsection{Supervised Fine-Tuning (SFT)}
Before preference alignment, we establish a base generation model to accelerate convergence and maintain training stability. 
To improve the generation quality associated with the highest musicality tier, we employ our automated music aesthetic evaluation framework as a strict data filter. 
Different from the broader top 5\% threshold used in pre-training, SFT retains only the top 0.5\% of songs per category based on average aesthetic scores. 
By conditioning the training data on the highest musicality-tier tag, we effectively narrow the output variance and ensure the model inherently favors top-tier musicality.

\subsubsection{Large-Scale Offline DPO for Controllability}
While SFT improves overall quality, the model may still suffer from lyrical hallucinations and unstable instruction following. 
To address this, we leverage large-scale offline DPO, which is highly efficient and cost-effective for enforcing controllability and stability through large-scale sampling and stringent paired filtering. 
Specifically, we generate 150,000 diverse lyrics via a large language model and randomly assign an audio prompt with its corresponding text description from a predefined dataset to each lyric. 
We then generate eight candidate samples per lyric under three distinct conditions: only the text description, only the audio prompt, and both the text description and the audio prompt.

The evaluation focuses on three core dimensions: lyric alignment, prompt consistency, and aesthetic score. To detect text hallucinations (e.g., word substitution, insertion, or deletion), we extract transcriptions using an Automatic Speech Recognition (ASR) model, convert them into phoneme sequences, and compute the edit distance against the target sequence, flagging samples with dense local anomalies. 
Prompt consistency is measured using the MuQ-MuLan \cite{zhu2025muq} model fine-tuned on text-described song data, while the aesthetic score is derived from the average dimension scores of our aesthetic framework. 

To prevent the model from learning from low-quality samples or noise, we discard the bottom 10\% of samples ranked by similarity and aesthetic scores. 
During preference pair construction, we prioritize controllability. For hallucinated samples, we select a hallucination-free sample generated under the same condition with higher musicality and no drop in similarity as the winning sample. 
For samples with low similarity, we pair them with a winning sample that exhibits a similarity improvement of at least 0.15, lacks hallucinations, and maintains musicality. 
Furthermore, if two samples exhibit a musicality gap of at least 0.5---a threshold corresponding to a 90\% expert agreement rate in our human evaluation comparison---we include this pair provided the winning sample is hallucination-free and shows equal or improved prompt similarity. 
This hierarchical pairing explicitly penalizes uncontrollable behaviors, reducing lyrical hallucinations and improving prompt consistency.

\subsubsection{Closed-Loop Semi-Online DPO for Musicality}
Following the offline stage, the model demonstrates stable instruction control. 
However, offline algorithms are bounded by the capabilities of the models used to construct the pre-defined datasets, and as training progresses, the model's policy gradually deviates from the static generation distribution, leading to a degradation in optimization efficacy. 
Inspired by recent advances in reinforcement learning \cite{lanchantin2025bridging}, we introduce a closed-loop semi-online DPO scheme. 
This approach reduces the limitations of offline algorithms and improves musicality while avoiding the computational costs of online sampling.

In this phase, we employ a periodically updated generator model to dynamically sample new data. 
To maintain a progressive alignment, the generator synchronizes its weights with the training model every 100 steps. 
We utilize a large language model to generate 160,000 new lyrics, and the generator produces 10 candidate samples for each input. 
Since base controllability is resolved in the previous stage, preference pair construction now focuses on musicality. 
For each input, we select the pair exhibiting the maximum musicality score gap, mandating that this gap be greater than 0.2---a threshold corresponding to a 73\% expert agreement rate in our human evaluation comparison. 
Crucially, to prevent regression, we enforce that the winning sample must remain hallucination-free and possess an instruction similarity score higher than the rejected sample.

By continuously generating fresh samples from the progressively improving policy, the model reduces its dependence on the static data distribution. 
Furthermore, to prevent reward hacking, we implement a hybrid validation protocol. 
This protocol combines human evaluation with multiple aesthetic assessment models to monitor the training progress and improve musicality while preserving stability.

\section{Experimental setup}
\label{sec:expsetup}
\paragraph{Dataset} 
LeVo 2 is trained on a large-scale music dataset comprising approximately 500,000 hours of song audio. 
To address the challenge of missing annotations, we adopt the automatic data processing pipeline proposed by SongPrep \cite{tan2025songprep} to conduct music segment analysis (e.g., verse, chorus), lyric recognition, and timestamp extraction. 
Furthermore, to obtain open-vocabulary tags, we utilize Qwen2-Audio \cite{chu2024qwen2} to annotate the training data.

\paragraph{Model setup}
Our proposed LeLM comprises approximately 4B parameters in total. 
Specifically, the Mixed Semantic LM utilizes a 36-layer Transformer with a hidden size of 2048, while the Track-Specific LM employs a 12-layer Transformer. 
For the MuEncoder model that generates mixed tokens and dual-track tokens, we utilize the pre-trained weights from MuCodec \cite{xu2024mucodec} with 300M parameters and a frame rate of 25 Hz.
Our diffusion model contains approximately 700M parameters to convert tokens into high-quality waveforms.
We replicate the VAE used in Stable Audio with 150M parameters.
Additionally, for lyrics and text descriptions, LeVo 2 employs a byte pair encoding (BPE)-based Qwen2 tokenizer \cite{yang2024qwen2} to process raw text. 

\paragraph{Training and Inference Details}
During the training of LeLM, we utilize 64 NVIDIA H20 GPUs with a batch size of 2 per GPU. 
The training process follows our three-stage paradigm: 600,000 steps for pre-training; followed by a decoupled post-training phase consisting of 5,000 steps for SFT, 5,000 steps for large-scale offline DPO, and 5,000 steps for closed-loop semi-online DPO; and finally, 200,000 steps for aesthetics-conditioned modular extension.
During the inference phase, we employ top-k sampling with k set to 5000 and a temperature of 0.8. 
Furthermore, to guide the model toward high-quality generation, we apply the musicality-aware CFG with a scale factor of 1.5.

\paragraph{Subjective Evaluations}
Considering that musical aesthetics and sound quality are difficult to evaluate accurately using automated models, we rely on expert subjective evaluations for these dimensions. 
We designed a six-dimensional Mean Opinion Score (MOS) listening test. It consists of four subjective artistic metrics: Overall Musicality (OVL), Melody (MEL), Arrangement (ARR), and Structure (STR); alongside two acoustic metrics: Vocal Sound Quality (SQ-V) and Accompaniment Sound Quality (SQ-A). We asked 20 music professionals to evaluate 100 Chinese songs and 100 English songs generated by each model, rating each dimension on a fine-grained scale from 1 to 10. To guarantee objectivity, none of the evaluators were involved in the model's development or training. During the test, participants were asked to listen to complete full-length songs generated by different models conditioned on the same input. This comparative setup ensures the resulting scores accurately reflect the relative performance differences between the systems. 

\paragraph{Objective Evaluations}
For the objective evaluation, we focus on instruction-following capabilities. To measure lyric alignment, we utilize SongPerP for lyric recognition and compute the Phoneme Error Rate (PER). To evaluate text-description alignment, we leverage Gemini as an automated evaluator. It assesses the alignment between the generated audio and the provided text prompts regarding genre and emotional expression (scored on a 1-10 scale). Furthermore, it explicitly identifies the instruments within the audio to compute the Instrument Generation Accuracy.

\begin{table*}[t]\small
  \caption{Subjective and objective evaluation results of LeVo 2 and comparison systems for song generation. The best overall results are highlighted in \textbf{bold}, and the best results among open-source models are \underline{underlined}. For all MOS results, the 95\% confidence intervals are within $\pm 0.05$.}
  \label{tab:main_results}
  \centering
  \begin{tabular}{lccccccccccc}
        \toprule
        \multirow{2}{*}{\textbf{Model}} & \multicolumn{6}{c}{\textbf{Subjective Evaluation (MOS $\uparrow$)}} & \multicolumn{3}{c}{\textbf{Gemini Evaluation $\uparrow$}} & \multicolumn{1}{c}{\textbf{ASR}} \\
        \cmidrule(lr){2-7} \cmidrule(lr){8-10} \cmidrule(lr){11-11}
         & \textbf{OVL} & \textbf{MEL} & \textbf{ARR} & \textbf{SQ-A} & \textbf{SQ-V} & \textbf{STR}  & \textbf{Genre} & \textbf{Emotion} & \textbf{Inst. (\%)} & \textbf{PER (\%)$\downarrow$} \\
        \midrule
        \multicolumn{11}{c}{\textit{Commercial Systems (Closed-source)}} \\
        \midrule
        Suno v5 & \textbf{5.72} & \textbf{6.38} & \textbf{6.55} & 7.10 & \textbf{6.77} & 6.17 & 7.93 & 8.68 & \textbf{93.06} & 12.40 \\
        Mureka v8 & 5.69 & 6.31 & \textbf{6.55} & \textbf{7.11} & 6.72 & \textbf{6.19} & \textbf{8.14} & 8.69 & 92.78 & 9.96 \\
        MiniMax Music 2.5+ & 5.22 & 5.87 & 6.02 & 6.83 & 6.49 & 5.60 & 7.44 & \textbf{8.72} & \textbf{93.06} & \textbf{7.80} \\
        \midrule
        \multicolumn{11}{c}{\textit{Open-source Systems}} \\
        \midrule
        ACE-Step 1.5 & 4.76 & 5.71 & 5.82 & 6.10 & 5.62 & 5.70 & \underline{7.21} & 8.35 & \underline{89.17} & 19.35 \\
        HeartMuLa & 4.07 & 4.94 & 5.05 & 5.80 & 4.98 & 4.92 & 4.42 & 7.55 & 76.11 & 19.88 \\
        DiffRhythm 2 & 2.95 & 4.05 & 4.86 & 4.06 & 3.83 & 3.88 & 5.72 & 8.02 & 82.50 & 11.88 \\
        YuE & 3.05 & 4.12 & 4.17 & 4.23 & 3.89 & 3.82 & 5.45 & 7.69 & 80.28 & 20.09 \\
        LeVo & 3.71 & 4.47 & 4.71 & 5.87 & 4.90 & 4.52 & 4.85 & 7.58 & 73.96 & 13.55 \\
        LeVo 2 (Ours) & \underline{5.48} & \underline{6.12} & \underline{6.42} & \underline{7.10} & \underline{6.53} & \underline{6.11} & 6.15 & \textbf{\underline{8.72}} & 88.35 & \underline{8.55} \\
        \midrule
    \end{tabular}
\end{table*}

\paragraph{Comparison systems}
We conducted a comprehensive comparison between LeVo 2 and multiple systems.
We selected three leading industry systems for benchmarking: Suno v5 \cite{suno}, Mureka v8 \cite{mureka}, and MiniMax Music 2.5+ \cite{minimax}. 
It is important to note that due to the black-box nature of these closed-source models, our evaluation conducted in May 2026 reflects the performance of these systems at that specific time.
Furthermore, we selected five open-source academic systems for benchmarking: YuE \cite{yuan2025yue}, DiffRhythm 2  \cite{jiang2025diffrhythm}, ACE-Step 1.5 \cite{gong2026ace}, HeartMuLa \cite{yang2026heartmula}, and LeVo \cite{leilevo}.

\section{Results and analysis} \label{sec:exp}
\subsection{Comparison with the State-of-the-Art Systems}
To comprehensively evaluate LeVo 2, we compare it against three leading commercial systems (Suno v5, Mureka v8, and MiniMax Music 2.5+) and five state-of-the-art open-source models. The comprehensive subjective and objective results are summarized in Table \ref{tab:main_results}.
\paragraph{Subjective Results}
In the subjective evaluations, LeVo 2 outperforms all open-source baselines across all six evaluated dimensions, highlighting its advantages in overall musicality, melody, arrangement, structure, as well as the sound quality of both vocals and accompaniment. Furthermore, compared to LeVo, the performance of LeVo 2 exhibits a significant leap. This improvement demonstrates the effectiveness of our proposed training paradigm, alongside our data optimization and scaling strategies. Although industry leaders like Suno v5 and Mureka v8 still maintain an advantage in many metrics, LeVo 2 approaches their performance closely, particularly in Melody (6.12), Arrangement (6.42), Accompaniment Sound Quality (7.10), and Structure (6.11). Notably, LeVo 2 surpasses MiniMax Music 2.5+ across all subjective listening metrics, showing that open-source architectures can approach commercial standards in this benchmark.

\paragraph{Objective results}
In the objective evaluations, LeVo 2 demonstrates competitive instruction-following capabilities. It achieves the lowest Phoneme Error Rate (PER) of 8.55\% among all open-source systems, indicating highly accurate lyric alignment and a significant reduction in lyrical hallucinations. Moreover, LeVo 2 achieves an Emotion control score of 8.72, which is the highest among all evaluated systems, including commercial ones. While LeVo 2 demonstrates robust controllability, we observe that its scores in Genre (6.15) and Instrument (88.35\%) control are slightly lower than those of ACE-Step 1.5, though it still outperforms the remaining open-source baselines. We attribute this gap primarily to the disparity in data annotation resources. ACE-Step 1.5 utilized an extremely precise annotation pipeline (e.g., utilizing Gemini 2.5 Pro to directly annotate 5 million songs, with the remainder annotated by a distilled model). In contrast, the scale and precision of our automated annotations were smaller. Despite this constraint in annotation quality, LeVo 2 still delivers competitive instrument control and achieves leading emotional accuracy, further demonstrating the robust alignment capabilities of our decoupled post-training strategy.

\subsection{Effectiveness of the Three-Stage Training Paradigm}
\begin{table*}[t]\small
  \caption{Evaluation Results of LeVo 2's capabilities across different training stages. The best overall results are highlighted in \textbf{bold}. For all MOS results, the 95\% confidence intervals are within $\pm 0.05$.}
  \label{tab:trajectory}
  \centering
  \begin{tabular}{lccccccccccc}
        \toprule
        \multirow{2}{*}{\textbf{Training Stage}} & \multicolumn{6}{c}{\textbf{Subjective Evaluation (MOS $\uparrow$)}} & \multicolumn{3}{c}{\textbf{Gemini Evaluation $\uparrow$}} & \multicolumn{1}{c}{\textbf{ASR}} \\
        \cmidrule(lr){2-7} \cmidrule(lr){8-10} \cmidrule(lr){11-11}
         & \textbf{OVL} & \textbf{MEL} & \textbf{ARR} & \textbf{SQ-A} & \textbf{SQ-V} & \textbf{STR} & \textbf{Genre} & \textbf{Emotion} & \textbf{Inst. (\%)} & \textbf{PER (\%)$\downarrow$} \\
        \midrule
        Pre-training Only & 4.61 & 5.26 & 5.57 & 6.34 & 5.77 & 5.46 & 5.67 & 8.14 & 85.67 & 11.09 \\
        \quad + SFT & 5.02 & 5.62 & 5.97 & 6.67 & 6.17 & 5.74 & 5.65 & 8.11 & 85.37 & 10.59 \\
        \quad \quad + Offline DPO & 5.12 & 5.77 & 6.11 & 6.80 & 6.30 & 5.82 & 5.90 & 8.59 & 87.78 & 9.19 \\
        \quad \quad \quad + Semi-Online DPO & 5.37 & 6.03 & 6.34 & 6.91 & 6.32 & 6.00 & 5.91 & 8.62 & 87.53 & 9.22 \\
        \quad \quad \quad \quad + Modular Ext. (LeVo 2) & \textbf{5.48} & \textbf{6.12} & \textbf{6.42} & \textbf{7.10} & \textbf{6.53} & \textbf{6.11} & \textbf{6.15} & \textbf{8.72} & \textbf{88.35} & \textbf{8.55} \\
        \bottomrule
    \end{tabular}
\end{table*}
To validate the contribution of each training phase, we present the step-by-step capability evolution of our model in Table \ref{tab:trajectory}. Initially, due to the uneven quality of the training data, the pre-trained model exhibits lower performance in both overall listening quality and controllability. Subsequently, the SFT stage, which strictly filters data utilizing the aesthetic evaluation framework, effectively enhances the overall listening experience. Specifically, the Overall Musicality jumps from 4.61 to 5.02, Melody improves from 5.26 to 5.62, and Arrangement increases from 5.57 to 5.97, proving that SFT successfully establishes a high-quality generation baseline. Building upon this, the large-scale offline DPO significantly reduces the Phoneme Error Rate (PER) from 10.59\% to 9.19\%. Simultaneously, the Gemini Evaluation scores show improvements without any degradation in musicality metrics. This demonstrates that the model improves the ability to follow instructions without sacrificing fundamental musicality. Once controllability is secured, the closed-loop semi-online DPO stage brings a comprehensive improvement in subjective artistic metrics, enhancing overall musicality, melody, and arrangement, while maintaining controllability and sound quality. This validates our decoupled strategy: by achieving robust controllability first and subsequently optimizing musicality independently, we effectively avoid gradient conflicts in multi-preference alignment and maximize the model's performance upper bound. Finally, through the Modular Extension stage, the Track-Specific LM further models and optimizes the acoustic details of individual tracks. This yields a significant enhancement in both Vocal Sound Quality (SQ-V) and Accompaniment Sound Quality (SQ-A), accompanied by further improvements in subjective artistic metrics and lyric accuracy (with PER dropping to 8.55\%), perfectly concluding our three-stage training paradigm.
\subsection{Analysis of Decoupled Progressive Post-Training}
\begin{table*}[t]\small
  \caption{Analysis of different post-training strategies based on the SFT model. The best results in each column are highlighted in \textbf{bold}. For all MOS results, the 95\% confidence intervals are within $\pm 0.05$.}
  \label{tab:post_train_analysis}
  \centering
  \begin{tabular}{lcccccccccc}
        \toprule
        \multirow{2}{*}{\textbf{Post-Training Strategy}} & \multicolumn{6}{c}{\textbf{Subjective Evaluation (MOS $\uparrow$)}} & \multicolumn{3}{c}{\textbf{Gemini Evaluation $\uparrow$}} & \multicolumn{1}{c}{\textbf{ASR}} \\
        \cmidrule(lr){2-7} \cmidrule(lr){8-10} \cmidrule(lr){11-11}
         & \textbf{OVL} & \textbf{MEL} & \textbf{ARR} & \textbf{SQ-A} & \textbf{SQ-V} & \textbf{STR} & \textbf{Genre} & \textbf{Emotion} & \textbf{Inst. (\%)} & \textbf{PER (\%)$\downarrow$} \\
        \midrule
        Pre-training + SFT & 5.02 & 5.62 & 5.97 & 6.67 & 6.17 & 5.74 & 5.65 & 8.11 & 85.37 & 10.59 \\
        \midrule
        \multicolumn{11}{c}{\textit{Single-Dimension Optimization}} \\
        \midrule
        Offline DPO (Musicality) & 5.23 & 5.86 & 6.15 & 6.82 & 6.18 & 5.87 & 5.59 & 8.17 & 86.18 & 11.41 \\
        Offline DPO (Lyrics Alignment) & 4.77 & 5.64 & 5.96 & 6.55 & 5.86 & 5.53 & 5.54 & 8.12 & 87.12 & \textbf{8.98} \\
        Offline DPO (Prompt Consistency)& 4.71 & 5.56 & 5.86 & 6.52 & 5.85 & 5.49 & \textbf{6.12} & \textbf{8.72} & \textbf{88.19} & 10.63 \\
        \midrule
        \multicolumn{11}{c}{\textit{Multi-Dimensional Optimization Baselines}} \\
        \midrule
        Mixed Training & 4.91 & 5.73 & 6.01 & 6.71 & 5.93 & 5.58 & 5.70 & 8.36 & 87.16 & 10.13 \\
        Interpolation & 5.17 & 5.80 & 6.13 & 6.77 & 6.21 & 5.90 & 5.75 & 8.46 & 87.58 & 9.93 \\
        \midrule
        \multicolumn{11}{c}{\textit{Ablations of Progressive Strategy}} \\
        \midrule
        Semi-Online DPO & 5.29 & 5.96 & 6.26 & 6.86 & 6.23 & \textbf{6.00} & 5.23 & 7.98 & 86.02 & 11.68 \\
        Offline DPO + Offline DPO (Mus.) & 5.21 & 5.87 & 6.21 & 6.78 & \textbf{6.36} & 5.94 & 5.87 & 8.60 & 87.62 & 9.42 \\
        Ours (Offline DPO + Semi DPO) & \textbf{5.37} & \textbf{6.03} & \textbf{6.34} & \textbf{6.91} & 6.32 & \textbf{6.00} & 5.91 & 8.62 & 87.53 & 9.22 \\
        \bottomrule
    \end{tabular}
\end{table*}
To validate the effectiveness of our proposed decoupled progressive post-training strategy, we compare it against various alignment strategies based on the SFT model, as shown in Table \ref{tab:post_train_analysis}. 

First, single-dimension optimization suffers from trade-off limitations. While optimizing for Lyrics Alignment achieves the lowest PER (8.98\%), or optimizing for Prompt Consistency yields the highest Genre (6.12) and Emotion (8.72) scores, these models experience substantial degradation in overall musicality (with OVL dropping to 4.77 and 4.71, respectively). Conversely, optimizing for musicality reduces the model's controllability. 

To address this issue, multi-dimensional optimization baselines (such as mixed training in HeartMuLa \cite{yang2026heartmula} and interpolation in LeVo \cite{leilevo}) attempt to balance multiple objectives simultaneously. However, due to gradient conflicts between objective controllability and subjective musicality, these methods fall into an averaging effect. For instance, their OVL scores stagnate at 4.91 and 5.17, and their controllability is lower compared to our proposed method. 

The proposed progressive strategy addresses these issues through decoupling. By first employing large-scale offline DPO to improve controllability to a stable level, and subsequently utilizing closed-loop semi-online DPO to further expand the upper bound of musicality, it achieves the optimal balance. It maximizes the overall musicality (highest OVL 5.37, MEL 6.03, and ARR 6.34) while preserving robust instruction-following capabilities (PER 9.22\%). 

Furthermore, ablation studies on our progressive strategy confirm this design. If the closed-loop Semi-Online DPO is applied directly, it leads to a decline in control capabilities. Conversely, replacing the final stage with another offline phase (offline DPO + offline DPO) maintains stability but is constrained by the static offline data, resulting in an OVL of 5.21.
\subsection{Ablation Studies}
\begin{table*}[t]\small
  \caption{Ablation studies on scaling, aesthetics-guided strategies, and architectural designs. For all MOS results, the 95\% confidence intervals are within \(\pm\) 0.05.}
  \label{tab:ablation_studies}
  \centering
  \begin{tabular}{lcccccccccc}
        \toprule
        \multirow{2}{*}{\textbf{Models}} & \multicolumn{6}{c}{\textbf{Subjective Evaluation (MOS $\uparrow$)}} & \multicolumn{3}{c}{\textbf{Gemini Evaluation $\uparrow$}} & \multicolumn{1}{c}{\textbf{ASR}} \\
        \cmidrule(lr){2-7} \cmidrule(lr){8-10} \cmidrule(lr){11-11}
         & \textbf{OVL} & \textbf{MEL} & \textbf{ARR} & \textbf{SQ-A} & \textbf{SQ-V} & \textbf{STR} & \textbf{Genre} & \textbf{Emotion} & \textbf{Inst. (\%)} & \textbf{PER (\%)$\downarrow$} \\
        \midrule
        LeVo 2 & \textbf{5.48} & \textbf{6.12} & \textbf{6.42} & \textbf{7.10} & \textbf{6.53} & \textbf{6.11} & 6.15 & \textbf{8.72} & 88.35 & 8.55 \\
        \midrule
        \multicolumn{11}{c}{\textit{Scaling \& Data}} \\
        \midrule
        \quad \textit{w/o} Model Scaling & 5.30 & 5.99 & 6.24 & 7.02 & 6.45 & 6.02 & 6.10 & 8.48 & 86.50 & 10.91 \\
        \quad \textit{w/o} Data Scaling & 5.27 & 5.92 & 6.14 & 6.92 & 6.31 & 5.87 & 5.32 & 8.42 & 81.03 & 14.29 \\
        \quad \textit{w/o} Pure Instrumental Data & 5.32 & 5.94 & 6.22 & 6.81 & 6.47 & 6.05 & 5.38 & 8.41 & 82.93 & \textbf{7.98} \\
        \midrule
        \multicolumn{11}{c}{\textit{Aesthetics-Guided Strategy}} \\
        \midrule
        \quad \textit{w/o} Musicality-Aware CFG & 5.33 & 5.98 & 6.18 & 6.90 & 6.48 & 5.90 & 5.91 & 8.60 & 86.45 & 9.35 \\
        \quad \textit{w/o} Aesthetics Guidance & 4.92 & 5.59 & 5.48 & 6.41 & 5.86 & 5.42 & \textbf{6.24} & 8.63 & \textbf{89.02} & 9.82 \\
        \midrule
        \multicolumn{11}{c}{\textit{Architecture Design}} \\
        \midrule
        \quad \textit{w/o} Delay Pattern & 4.65 & 4.93 & 5.01 & 5.91 & 5.32 & 4.79 & 5.37 & 7.39 & 72.09 & 47.10 \\
        \quad \textit{w/o} Track-Specific LM & 4.89 & 5.32 & 5.47 & 5.69 & 4.92 & 3.51 & 4.89 & 6.73 & 61.76 & 22.41 \\
        \quad \textit{w/o} Acoustic Augmentation & 5.43 & 6.12 & 6.41 & 6.83 & 6.31 & 6.01 & 6.02 & 8.61 & 88.51 & 9.03 \\
        \bottomrule
    \end{tabular}
\end{table*}
To further investigate the impact of our designs, we conduct ablation studies focusing on scaling, aesthetics-guided strategies, and architecture design. The results are detailed in Table \ref{tab:ablation_studies}.
\paragraph{Scaling and Data Strategy}
To explore the scaling strategies in LeVo 2, we first trained a smaller LeLM with approximately 2B parameters (\textit{w/o} Model Scaling) and the standard LeLM based on only 200,000 hours of songs (\textit{w/o} Data Scaling). The results indicate that both ablations lead to a decline across almost all subjective and objective metrics. Most notably, the reduction in data scale results in a significant decline in controllability. Furthermore, when we remove pure instrumental music from the training data (\textit{w/o} Pure Instrumental Data), the model suffers a noticeable drop in Accompaniment Sound Quality and Genre alignment. This demonstrates that pure instrumental data is crucial for modeling diverse song genres and high-fidelity accompaniments.

\paragraph{Aesthetics-Guided Strategy}
To validate the effectiveness of the aesthetics-guided strategy, we conducted two ablation experiments. First, removing the aesthetic tier tags during pre-training and modular extension (\textit{w/o} Aesthetics Guidance) leads to a substantial decrease in subjective metrics. Notably, its genre and instrument control scores show a slight increase. This indicates that aesthetic tier tags effectively help the model learn musical priors, but the aesthetic evaluation framework still exerts a preference bias on genre and instrument alignment. Furthermore, removing the musicality-aware CFG causes a drop in subjective metrics, confirming its necessity during inference.

\paragraph{Architecture Design}
We further conducted ablations on the techniques employed in our model. We observe that removing either the Track-Specific LM or the delay pattern between the dual-track tokens and mixed tokens leads to a significant decline across all metrics. This validates the necessity of our hierarchical architecture and the importance of providing semantic context to the Track-Specific LM. Furthermore, removing the acoustic augmentation strategy during the modular extension stage decreases both Vocal and Accompaniment Sound Quality. This confirms that this intentional perturbation successfully forces the Track-Specific LM to learn and restore fine-grained acoustic nuances.

\section{Discussion and Ethics Statement}
\label{sec:discussion_ethics}

Despite the strong empirical results, song generation remains a challenging task. First, although LeVo 2 outperforms all academic baselines in our evaluation, its audio quality is still constrained by the variability and scale of the training data. As a result, a gap remains between LeVo 2 and the strongest proprietary systems. Second, the scarcity and high cost of annotated music data lead us to rely heavily on open-source models for pseudo-labeling, which inevitably affects the model's instruction-following ability. On the one hand, because text descriptions are generated using Qwen2-Audio, the diversity and richness of effective prompts are limited. On the other hand, accumulated errors in lyric extraction, structure recognition, and music captioning introduce noise into the supervision process and reduce the accuracy of instruction alignment. These factors explain why, although our progressive post-training strategy improves both controllability and musicality, LeVo 2 still shows a controllability gap compared with top proprietary systems.

We fully recognize the potential ethical risks associated with music generation models. All models and training data used in this work are intended only for academic research purposes. We respect the intellectual property rights of original artists and content creators, and have taken reasonable measures to avoid using copyrighted materials without appropriate authorization.

\section{Conclusion}
In this paper, we present LeVo 2, a novel hybrid LLM-Diffusion framework designed for stable and melodious full-length song generation. We introduce a hierarchical representation modeling architecture to decouple the generation process into a Mixed Semantic LM for global semantic orchestration and a Track-Specific LM for parallel dual-track acoustic refinement. Furthermore, to overcome the challenges posed by the scarcity of high-quality data and gradient conflicts in multi-preference alignment, we propose an aesthetics-guided three-stage training paradigm featuring a decoupled progressive post-training strategy. By sequentially integrating Supervised Fine-Tuning (SFT), large-scale offline DPO, and closed-loop semi-online DPO, our framework successfully mitigates lyrical hallucinations and maximizes artistic expressiveness beyond the limitations of static datasets. Extensive subjective and objective evaluations demonstrate that LeVo 2 consistently outperforms state-of-the-art open-source models and narrows the gap to proprietary systems. We release our model weights and code to facilitate future advancements in automated high-fidelity music generation.

\bibliography{refs}
\bibliographystyle{ieeetr}

\end{document}